# The induced subgraph $K_{2,3}$
# in a non-Hamiltonian graphs

Heping Jiang [0000-0001-5589-808X]

jjhpjhp@gmail.com

## Abstract

A graph $G$ is a tuple $(V, E)$, where $V$ is the vertex set, $E$ is the edge set. A reduced graph is a graph of deleting non-Hamiltonian edges and smoothing out the redundant vertices of degree 2 on an edge except for leaving only one vertex of degree 2. We denote by $I$ a set of cycles only jointed by inside vertices. $|I|$ is the number of sets $I$ in a graph. We use a norm graph to denote a reduced graph of $|I|=1$. $g$ is a subgraph obtained by deleting all removable cycles from a basis of a norm graph. In this paper, we show that a norm graph $G$ is non-Hamiltonian, if and only if, $g$ and $K_{2,3}$ are homeomorphic.

**Keywords:**   Hamilton graph, set $I$, norm graph, subgraph $g$, $K_{2,3}$ set

**Mathematics Subject Classification 2010:** 05C45

## 1   Introduction

Graphs considered in this paper are finite, undirected, and simple connected graphs. A graph $G$ is a tuple $(V, E)$, where V is the vertex set, E is the edge set. A reduced graph is a graph of deleting non-Hamiltonian edges and smoothing out the redundant vertices of degree 2 on an edge except for leaving only one vertex of degree 2. We denote by $I$ a set of cycles only jointed by inside vertices. $|I|$ is the number of sets $I$ in a graph. We use a norm graph to denote a reduced graph of $|I|=1$. $g$ is a subgraph obtained by deleting all removable cycles from a basis of a norm graph.

In this paper, by observing the 2-common $(v, 0)$ combination [1] in a basis of the cycle spaces of norm graphs, we relate a bipartite graph $K_{2,3}$ to norm non-Hamilton graphs and show that the existence of a unique induced subgraph $K_{2,3}$ is a sufficient and necessary condition of norm non-Hamilton graphs.

## 2 Preliminaries

### Basic definitions

A graph $K_{m,n}$ is a bipartite graph if its vertex set can be partitioned into two subsets $m$ and $n$ so that every edge has one end in $m$ and one end in $n$. The subdivision of an edge e with endpoints $\{u, v\}$ is an operation that adds one new vertex w and an edge set replacing e by two new edges $\{u, w\}$ and $\{w, v\}$. The reverse operation is called smoothing that deletes a vertex w and the two edges $\{u, w\}$ and $\{w, v\}$, and replaces with an edge e (Only the vertices of degree 2 can be smoothed). We use $G_1 \approx G_2$ to define two graphs $G_1$ and $G_2$ are homeomorphic such that $G_1$ can be transformed into $G_2$ mutually via a finite sequence of subdivisions and smoothes through vertices of degree 2. An induced subgraph is a subgraph obtained by deleting a vertex or set of vertices. A vertex $v$ is incident with an edge $e$ if $v \in e$. A boundary vertex is a vertex that only has two edges of $R=1$ in its incident edges. A vertex is called inside if it is neither a boundary vertex nor a cut point.

### Some new notations and terminologies

A reduced graph is a graph of deleting non-Hamiltonian edges and smoothing out the redundant vertices of degree 2 except for leaving only one vertex of degree 2 on an edge. Let $I$ be a set of cycles only jointed by inside vertices adjacent each other. $|I|$ is the number of set $I$ in a graph. We use a norm graph to denote a reduced graph of $|I|=1$. $g$ is a subgraph derived by deleting all removable cycles from a basis of a norm graph.

There are various cases of reduced graphs of $|I| \geq 2$. However, for a graph of $|I| \geq 2$, if the Hamiltoncity for every subset of $I$ is given, then we can easily identify the Hamiltoncity of the whole graph. Thus, to determine the Hamiltoncity of graphs, without loss of generality, it is enough to consider the case of $|\boldsymbol{I}|=2$ in discussion. Apparently, in the case of $|\boldsymbol{I}|=2$, there only have three subcases. See the Figure 2.1. And clearly, we can easily determine the Hamiltoncity of these three graphs by identifying the Hamiltoncity of $\boldsymbol{I}_1$ and $\boldsymbol{I}_2$ respectively. For simplicity, in the following contents in this paper, we only discuss the reduced graphs of $|\boldsymbol{I}|=1$, the norm graphs, unless specified otherwise.

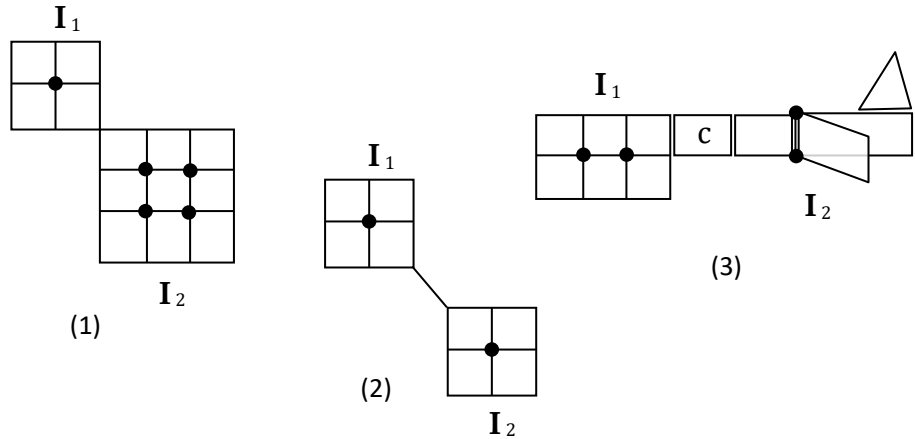

**Figure 2.1**   In graph (1) $\mathbf{I}_1$ connects $\mathbf{I}_2$ by a cut point, in graph (2) $\mathbf{I}_1$ connects $\mathbf{I}_2$ by a bridge, and in graph (3) $\mathbf{I}_1$ connects $\mathbf{I}_2$ by a cycle c. All the bold black points are inside vertices.

For the Hamiltoncity of $K_{2,3}$, we have

**Proposition 2.1** $K_{2,3}$ *is a non-Hamilton graph with minimum number of graphic elements.*

*Proof.*   For a bipartite graph, by Rule 2.1, $K_{2,3}$ is a non-Hamilton graph. Since $K_{2,3}$ is a norm graph, then deleting any graphic element from $K_{2,3}$ will produce $|\boldsymbol{I}|\neq1$. Thus, $K_{2,3}$ is a graph with minimum number of graphic elements. Hence, the statement holds. ∎

We write $\mathcal{g}\cong K_{2,3}$ to define $\mathcal{g}\approx K_{2,3}$ such that there has only one subgraph $K_{2,3}$ induced from $\mathcal{g}$.

Definitions and terminologies not mentioned in this paper please refer to [2-4].

## 3  Main results and proofs

Let $G$ be a norm graph and $\mathcal{g} \subseteq G$. We have the following results.

**Lemma 3.1**  *G is non-Hamiltonian* $\Rightarrow \mathcal{g} \cong K_{2,3}$.

*Proof.*    We first prove $\mathcal{g} \approx K_{2,3}$. From $G$ is non-Hamiltonian, we know first that $G$ has no solution, by Lemma 1.2 in [1], $\mathcal{g}$ is neither a 2-common $(v, e)$ combination nor a 2-common $(v, 0)$ combination. According to Rules in [5], we derive $\mathcal{g} \approx K_{2,3}$. Secondly, consider $G$ is a non-Hamiltonian graph but has a solution, which means $\mathcal{g}$ is a 2-common $(v, 0)$ combination. By lemma 3.1 [5], $|C_k| \neq 0 \Leftrightarrow |P| \geq 3$, according Proposition 2.1, we have $\mathcal{g} \approx K_{2,3}$.

Secondly, we prove that only one $K_{2,3}$ can be induced from $\mathcal{g}$. By the Hamiltoncity described in Rules in [5], without loss of generality, we suppose that there have two induced subgraphs $K_{2,3}$ in a basis of $G$. Then, there have three sub-cases as followings in Figure 3.1.

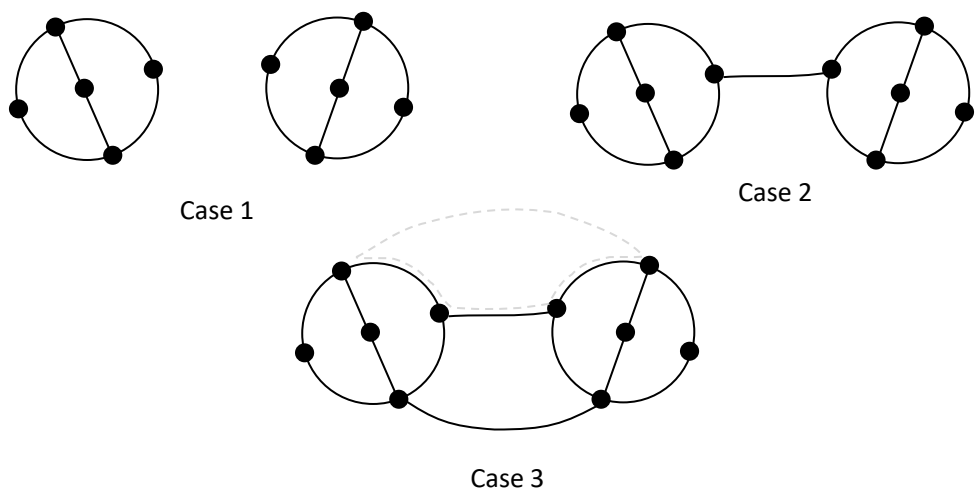

**Figure 3.1** Three sub-cases including two $K_{2,3}$ subgraphs

For a norm graph, by the definition of $\mathcal{g}$, it is clearly that Case 1 does not exist.

In Case 2, we have three possible choices to construct the last removable cycle in generating $\mathcal{g}$ from a basis of $G$, which marked as the dotted edges $a$, $b$, and $c$ in Figure 3.2. But, whatever the edge $e$ combines with anyone of the dotted edges, there are no removable cycles in the combination. So there have two $K_{2,3}$ induced from $G$, then we obtain $|I| \neq 1$. This contradicts to that $G$ is a norm graph. Thus, the assumption of existence of two $K_{2,3}$ does not hold in Case 2.

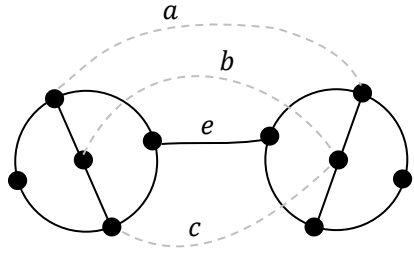

**Figure 3.2**

In Case 3, we have five possible choices for constructing the last one removable cycle in generating $g$ from a basis of $G$, see the dotted arcs in Figure 3.3.

For graph (1) and graph (2), we first let the cycle including edges $a$ and $e_1$ or the cycle including edges $a$ and $e_2$ be a removable cycle respectively. Whatever the dotted edge $a$ combines with either of the edges $e_1$ and $e_2$, there will be two $K_{2,3}$ induced from $G$, that is same to the result of Case 2. And Secondly, we consider the cycle including edges $e_1$ and $e_2$ as a removable cycle. After deleting it, we obtain a 2-common ($v$, $e$) combination. It contradicts to that $G$ is non-Hamiltonian. Therefore, the assumption of existence of two $K_{2,3}$ does not hold in graph (1) and graph (2).

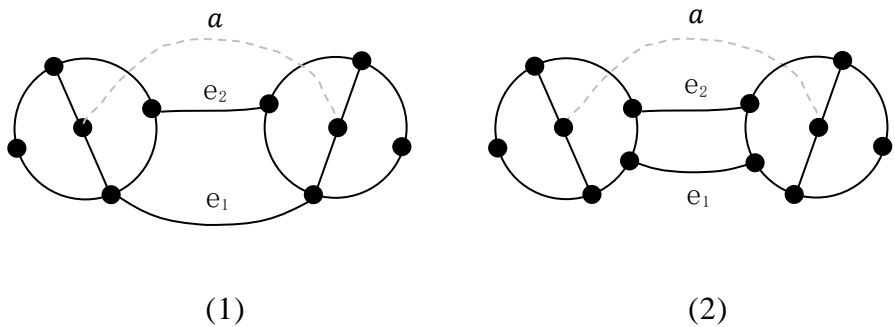

(1)                    (2)

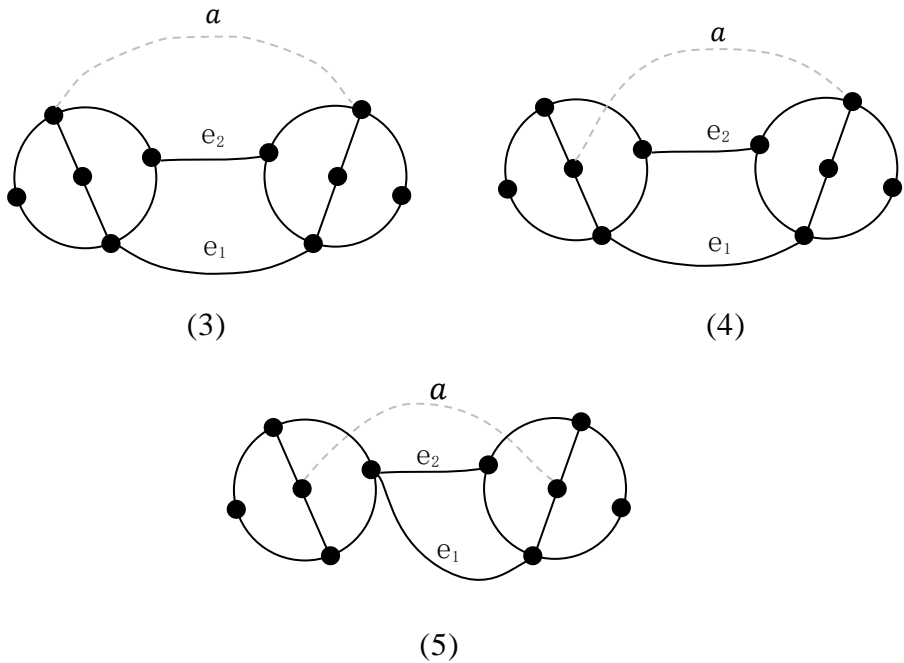

**Figure 3.3** Five possible choices in Case 3

For graph (3) in Case 3, by the definition of set $\boldsymbol{I}$, obviously there exists $|\boldsymbol{I}| \geq 2$. Hence, the assumption of existence of two $K_{2,3}$ does not hold in graph (3).

For graph (4) in Case 3, if we let the cycle including edges $a$ and $e_1$ or the cycle including edges $a$ and $e_2$ be a removable cycle respectively, then the left graph will include two subgraphs $K_{2,3}$, which indicates $|\boldsymbol{I}| \neq 1$. It contradicts to that $G$ is a norm graph. Otherwise, if we let the cycle including edges $e_1$ and $e_2$ be a removable cycle, then we obtain a cycle set including one subgraph $K_{2,3}$. Hence, the assumption of existence of two $K_{2,3}$ does not hold in graph (4).

For graph (5) in Case 3, there only have two subgraphs obtained by deleting a removable cycle from a basis of $G$. The first obtained subgraph is induced by deleting a removable cycle including the edge $e_1$ such that the subgraph has no the edge $e_1$. The result is same to that of graph (1) and graph (2), which is a 2-common ($v$, $e$) cycle set. It contradicts to that $G$ is non-Hamiltonian. The second obtained subgraph is induced by deleting a removable cycle including the edge $e_2$ such that the subgraph has no the edge $e_2$. This result is same to the case of graph (4), in which the cycle including edges $e_1$ and $e_2$ is a removable cycle. And so, we obtain a cycle set including one subgraph $K_{2,3}$. Hence, the assumption of existence of two $K_{2,3}$ does not hold in graph (5).

Thus, if $g \approx K_{2,3}$, the statement that only one subgraph $K_{2,3}$ can be induced from $g$ holds.

We complete the proof. ∎

**Lemma 3.2**    *G is Hamiltonian $\Rightarrow g \approx C_3$.*

*Proof.* $G$ is Hamilton graph, by Lemma 3.3 in [5], we have $|C_k|=0$. By Lemma 1.3 in [1], for $g \subseteq G$, we have $g$ is a 2-common ($v$, $e$) cycle set. Note that the sum of the cycle set is a Hamilton cycle in $G$. By the definition of graph homeomorphism, we can easily derive that $g \approx C_3$. ∎

**Theorem 3.3**    *G is non-Hamiltonian $\Longleftrightarrow g \cong K_{2,3}$.*

*Proof.*    By Lemma 3.1, we have that $G$ is non-Hamiltonian $\Rightarrow g \cong K_{2,3}$. By Lemma 3.2, $G$ is Hamiltonian $\Rightarrow g \approx C_3$, which equal to $g \ncong K_{2,3}$. Hence, the statement holds. ∎

## Acknowledgements

The author thanks anonymous reviewers for their very careful review and multiple comments that helped improve the presentation.